\documentclass[prl,twocolumn,showpacs,amsmath,amssymb,superscriptaddress]{revtex4-1}

\usepackage{bm}
\usepackage{graphicx}
\usepackage[usenames,dvipsnames,svgnames,table]{xcolor}
\usepackage[colorlinks=true,linkcolor=RoyalBlue,citecolor=RoyalBlue]{hyperref}

\begin{document}

\title{Gate-controllable magneto-optic Kerr effect in layered collinear antiferromagnets}

\author{Nikhil Sivadas}
\affiliation{Department of Physics, Carnegie Mellon University, Pittsburgh, Pennsylvania 15213, USA}

\author{Satoshi Okamoto}
\affiliation{Materials Science and Technology Division, Oak Ridge National Laboratory, Oak Ridge, Tennessee 37831, USA}

\author{Di Xiao}
\affiliation{Department of Physics, Carnegie Mellon University, Pittsburgh, Pennsylvania 15213, USA}
\date{\today}

\begin{abstract} 

Using symmetry arguments and a tight-binding model, we show that for layered collinear antiferromagnets, magneto-optic effects can be generated and manipulated by controlling crystal symmetries through a gate voltage. This provides a promising route for electric field manipulation of the magneto-optic effects without modifying the underlying magnetic structure.  We further demonstrate the gate control of magneto-optic Kerr effect (MOKE) in bilayer MnPSe$_3$ using first-principles calculations.  The field-induced inversion symmetry breaking effect leads to gate-controllable MOKE whose direction of rotation can be switched by the reversal of the gate voltage. 

\end{abstract}
\maketitle

Magneto-optic effects are one of the defining features of time-reversal ($\mathcal T)$ symmetry breaking in matter.  Usually, the $\mathcal T$ symmetry is broken either by an external magnetic field, or by the spontaneous appearance of a macroscopic magnetization such as in ferromagnets.  Similar to their ferromagnetic counterparts, the $\mathcal T$ symmetry is also broken in antiferromagnets.  However, because of their vanishing net magnetization one would naively expect an absence of magneto-optic effects in antiferromagnets.  This assumption has been recently challenged by the theoretical demonstration of a rather large magneto-optic Kerr effect (MOKE) in certain non-collinear antiferromagnets with zero net magnetization~\cite{Feng15p144426}.  This effect is closely related to the anomalous Hall effect predicted in the same class of materials~\cite{Chen14p017205,kubler2014}, both of which are dictated by the absence of certain crystal symmetries.  The appearance of magneto-optic effects in antiferromagnets is of intrinsic interest, since it would allow direct detection of the magnetic order and therefore could be useful for antiferromagnets-based memory devices~\cite{jungwirth2016}.

While non-collinear antiferromagnets have been the focus of recent interest~\cite{Feng15p144426,Chen14p017205,kubler2014}, in this Letter we show that magneto-optic effects can also exist in the more commonly available collinear antiferromagnets.  We start by analyzing the general symmetry requirements for magneto-optic effects, and demonstrate the symmetry principles by constructing a tight-binding model with a collinear N\'eel type order.  We show that, contrary to the general belief, lifting the spin degeneracy of the energy bands is not a sufficient condition to generate magneto-optic effects; it is the crystal symmetry that actually controls these effects.  

Based on this understanding, we predict that a perpendicular electric field can be used to generate and control the MOKE in layered antiferromagnets using first-principles calculations.  Recent theoretical and experimental progress has identified several layered compounds as promising candidates to host magnetism in their thin-film limit~\cite{Casto15p041515,Sivadas15p235425,Williams15p144404,McGuire15p612,lin2016,Li13p3738}.  One of them is MnPSe$_3$, a semiconductor with collinear antiferromagnetic order within each layer.  We show that the field-induced inversion ($\mathcal{I}$) symmetry breaking in bilayer MnPSe$_3$ gives rise to a MOKE whose direction of rotation can be switched by the reversal of the gate voltage.  Our result indicates that layered antiferromagnets would provide a very promising platform to explore gate-controllable magneto-optic effects.  

\begin{figure}
\includegraphics[width=1.0\columnwidth]{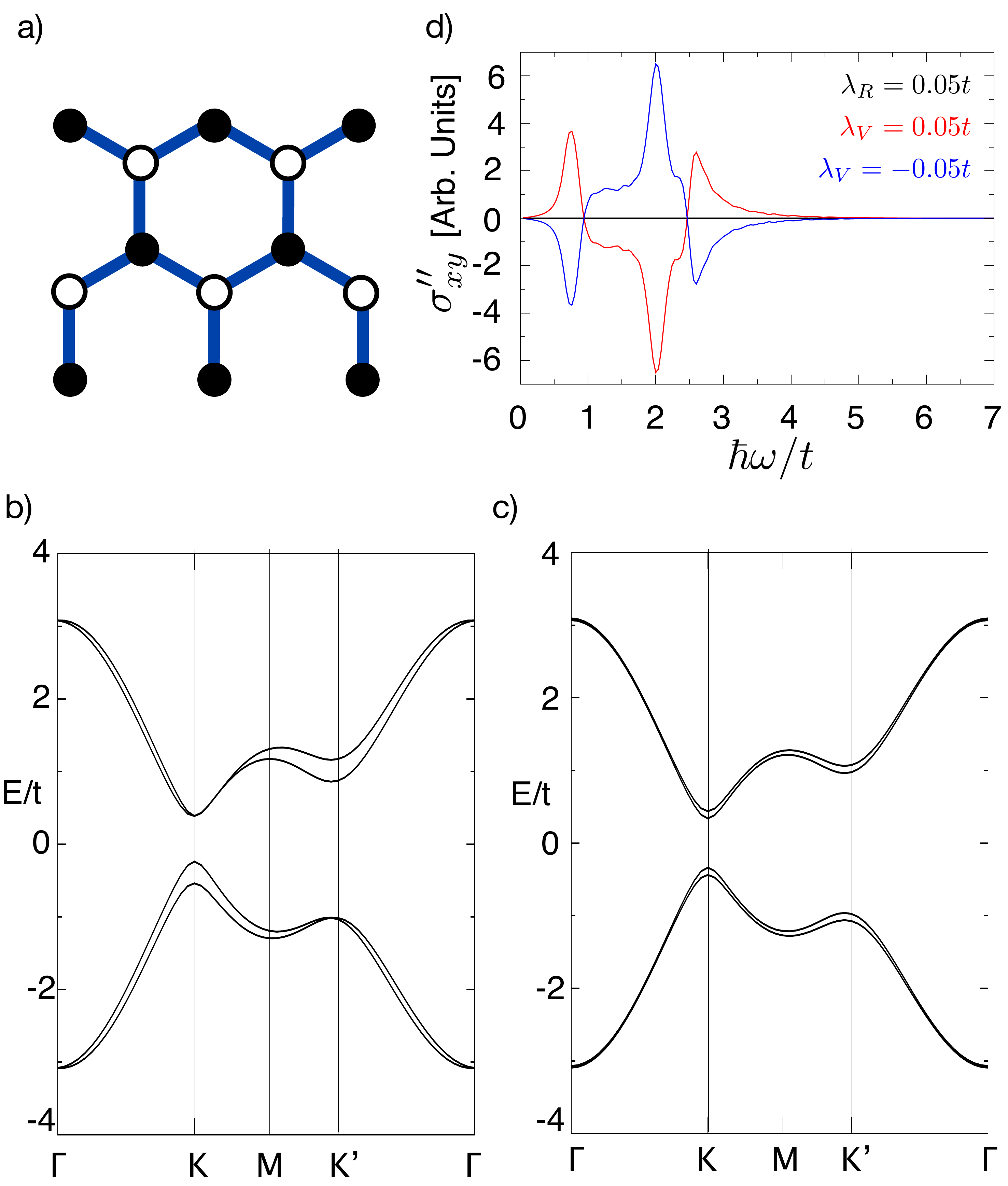}
\caption{\label{fig:TB} (Color online)  (a) Schematic of a honeycomb lattice with collinear N{\'e}el order.  Up (down) spins are represented by filled (open) circles.  The system possesses combined $\mathcal{TI}$ symmetry although both $\mathcal{T}$ and $\mathcal{I}$ symmetries are individually broken.  (b) and (c) Energy bands of the tight binding model with broken mirror symmetry ($\lambda_R = 0.05t$, $\lambda_V = 0$) and broken in-plane inversion symmetry ($\lambda_R = 0$, $\lambda_V = 0.05t$), respectively. In both cases, $\lambda_{SO} = 0.06t$ and $\lambda_M=0.7t$. The spin degeneracy of the bands is lifted in both cases. (d) The imaginary part of the optical Hall conductivity ($\sigma''_{xy}$) computed for $\lambda_R = 0.05t$ (black), $\lambda_V = 0.05t$ (red) and $\lambda_V = -0.05t$ (blue).  $\sigma''_{xy}$ is zero when only $\lambda_R$ is turned on and becomes non-zero when $\lambda_V \neq 0$.  As the sign of $\lambda_V$ is reversed so is $\sigma''_{xy}$.  The smearing parameter was set to $0.1t$.}
\end{figure}

As symmetries play an important role in magneto-optic effects~\cite{Eremenko89}, we begin our discussion with a general symmetry analysis. Magneto-optic effects are closely related to the AC Hall effect [see Eq.~\eqref{kerr} below], which refers to the appearance of a transverse AC current in response to an optical field in the longitudinal direction. Therefore, we can use the following pseudo-vector
\begin{equation} \label{j}
\bm n  = \bm j \times \bm E
\end{equation}
to characterize magneto-optic effects.  If the material possesses $\mathcal{T}$ symmetry, $\bm n$ is clearly constrained to be zero. Both ferromagnets and antiferromagnets break $\mathcal{T}$ symmetry. However, it is possible that the material might have a combined symmetry of $\mathcal T$ and some crystal symmetry $\mathcal{O}$, which can force $\bm n$ to be zero even if $\mathcal T$ symmetry is broken.  To elucidate this, consider an antiferromagnets with $\mathcal{TI}$ symmetry. One such example is shown in Fig.~\ref{fig:TB}(a).  Under $\mathcal{TI}$ symmetry, $\bm j$ is unaffected, whereas $\bm E$ changes sign.  It then follows from Eq.~(\ref{j}) that $\bm n$ changes sign under the $\mathcal{TI}$ symmetry operation.  This forces $\bm n$ to be zero and suppresses any magneto-optic effects. Using a similar analysis, it is straightforward to show that for two-dimensional systems both $\mathcal{TM}_z$ symmetry and $\mathcal{TC}_2$ symmetry also suppress magneto-optic effects, where $\mathcal M_z$ is the mirror reflection perpendicular to the $\bm j$-$\bm E$ plane, and $\mathcal{C}_2$ is the in-plane inversion symmetry.  Thus, by breaking these crystal symmetries, magneto-optic effects can be generated in antiferromagnets. This is the key to our gate controllable MOKE.

Armed with the above insight, we now consider a specific example, a honeycomb lattice with a collinear N{\'e}el type order, as shown in Fig.~\ref{fig:TB}(a).  The Hamiltonian is given by
\begin{equation}
H = t\sum_{\langle ij \rangle} c_i^{\dagger}c_j + i \lambda_{SO}\sum_{\langle\langle ij \rangle \rangle} \nu_{ij}  c_i^{\dagger} s^{z} c_j 
+ \sum_i (-1)^{i}\lambda_{M}c_i^{\dagger}s^{z}c_i \;.
\end{equation}
The first term is the nearest neighbor hopping.  The second term is the intrinsic spin-orbit coupling (SOC), which is needed for any magneto-optic effects.  Here, $\nu_{ij} = (2/\sqrt3)(\hat{\bm{d}}_{1} \times \hat{\bm{d}}_{2})_{z} = \pm1 $, where $ \hat{\bm{d}}_{1}$ and  $\hat{\bm{d}}_{2}$ are the unit vectors of the two bonds connecting site $i$ to $j$, and $s^{z}$ is the spin Pauli matrix.  Along with preserving the $\mathcal{M}_z$ symmetry,  this term also preserves both $\mathcal{T}$ and $\mathcal{I}$  symmetries.  The third term breaks $\mathcal{T}$ symmetry via a staggered Zeeman field, mimicking the N{\'e}el order with an out-of-plane easy axis.  We note that this term can be dynamically generated by local interactions, $\sum_i U n_{i, \uparrow}n_{i, \downarrow}$  \cite{Rachel2010,Hohenadler2011}. Within the mean-field approximation, $U$ and $\lambda_M$  are related by $\lambda_M = \frac{m}{2} U$ where $m = \langle n_{i, \uparrow} - n_{i, \downarrow} \rangle$ is the spontaneous magnetic moment. Thus, our results are also valid for interacting systems with robust magnetic ordering.  One can verify that the system is invariant under the $\mathcal{TI}$ symmetry.  This Hamiltonian is identical to the one proposed by Kane and Mele for the quantum spin Hall effect~\cite{Kane05p146802}, except the $\lambda_M$ term.  As we are interested in the properties of a topologically trivial antiferromagnetic insulator, we will work in the strong exchange limit where the band gap is dominated by $\lambda_M$ ($\lambda_M \gg 3\sqrt3\lambda_{SO}$).

To analyze the role of crystal symmetries, we add two symmetry breaking terms to the Hamiltonian
\begin{equation}
H' = i \lambda_R \sum_{\langle ij \rangle} c_i^{\dagger} (\bm{s} \times \hat{\bm{d}}_{ij})_{z} c_j   + \lambda_V\sum_i (-1)^i c_i^{\dagger}c_i \;.
\end{equation}
The Rashba SOC term ($\lambda_R$) breaks the $\mathcal{M}_z$ symmetry, and the staggered sublattice potential ($\lambda_V$) breaks the in-plane inversion symmetry.  Figure~\ref{fig:TB}(b) and (c) show the energy bands obtained for two representative cases where the $\mathcal{TI}$ symmetry is broken.  In cases I we switch on only the Rashba term ($\lambda_R \ne 0$), whereas in case II only the staggered sublattice potential is turned on ($\lambda_V \ne 0$).  It is clear that the effect of these $\mathcal{TI}$ symmetry breaking terms is to lift the spin degeneracy of the bands.  We also note that $K$ and $K'$ valleys are no longer degenerate.  This is not a consequence of $\mathcal{TI}$ symmetry breaking, and in fact, they remain non-degenerate even when the symmetry breaking terms are removed.  The breaking of the valley degeneracy arises from the interaction of the antiferromagnetic order and the intrinsic SOC~\cite{Li13p3738}.

Next, we calculated the optical Hall conductivity $\sigma_{xy}(\omega)$ using the Kubo-Greenwood formula~\cite{yao2004,Ebert96p1665}, 
\begin{equation}
\begin{split}
\sigma_{xy}(\omega) &= \hbar  e^2 \int  \frac{d^2k}{(2\pi)^2}\sum_{n \ne m} (f_{m\bm{k}}-f_{n\bm{k}} ) \\
&\quad \times\frac{ \text{Im} \langle\psi_{n\bm{k}} | v_x|\psi_{m\bm{k}}\rangle \langle \psi_{m\bm{k}} | v_y|\psi_{n\bm{k}}\rangle}{(\varepsilon_{m\bm{k}}-\varepsilon_{n\bm{k}})^2-(\hbar\omega+i\eta)^2}, 
\end{split}
\end{equation}
where $f_{m\bm k}$ is the Fermi-Dirac distribution function, $\varepsilon_{m\bm{k}}$ is the energy of the $m$th band,  $\hbar \omega$ is the photon energy, and $\eta$ is an adjustable smearing parameter with units of energy. Figure~\ref{fig:TB}(d) shows the imaginary part of $\sigma_{xy}$, denoted by $\sigma''_{xy}$.  Even though the bands are spin-split in both cases, we can see that $\sigma''_{xy}$ is identically zero for case I and is non-zero only for case II.  To understand this we further analyze the symmetry properties of the system.  We note that even though the system is invariant under $\mathcal{TI}$, the $\mathcal{TM}_z$ symmetry is already broken by the out-of-plane magnetic order.  In case I, although the Rashba term breaks $\mathcal{M}_z$ symmetry, the system still possesses $\mathcal{TC}_2$ symmetry. As we discussed earlier, it suppresses any magneto-optic effects.  This shows that even though the bands are spin-split, the underlying crystal symmetries can force the magneto-optic effects to vanish.  In case II, the staggered sublattice potential breaks both $\mathcal{TI}$ and $\mathcal{TC}_2$ symmetries, it therefore lifts all symmetry constraints on magneto-optic effects, making it non-zero.

In addition, we also find that upon the reversal of the staggered sublattice potential, $\sigma''_{xy}$ changes its sign. It can be verified that the process of reversing the sign of the staggered sublattice potential is equivalent to switching the sublattices and reversing the spins. This operation is nothing but the $\mathcal{TI}$ symmetry operation.  However, we have already discussed that $\mathcal{TI}$ symmetry operation reverses the sign of $\sigma_{xy}$, which is indeed what we find.  On the other hand, if natural birefringence also exist in the system, their contribution would not flip sign upon the reversal of the sublattice potential.  This property can be used to distinguish between magneto-optical effects and natural birefringence.

\begin{figure}
\includegraphics[width=1.0\columnwidth]{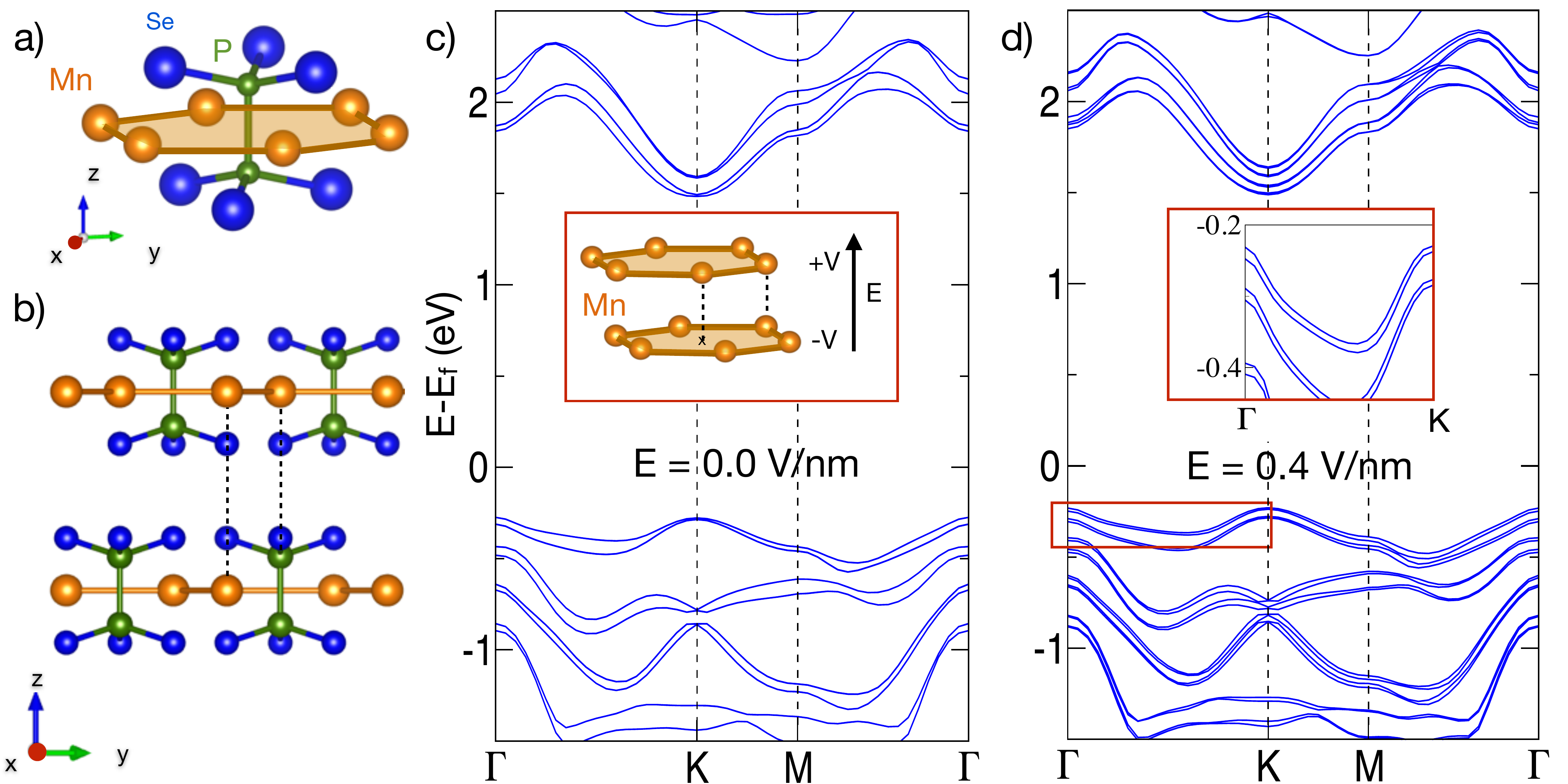}
\caption{\label{fig:Fig2} (Color online) (a) The crystal structure of monolayers of MnPSe$_3$. The transition metal Mn atoms form a honeycomb structure with P$_2$Se$_6$ ligand occupying the center of the honeycomb. (b) The side view of the crystal structure of bilayer MnPSe$_3$.  The crystal structure is drawn using VESTA~\cite{Momma11p1272}. (c) The band structure of the bilayer MnPSe$_3$ in the absence of an electric field. The insert shows the Mn atoms in the bilayer.  (d) The band structure of the bilayer MnPSe$_3$ in the presence of an electric field (0.4~V/nm) along the $z$-direction.  The insert shows the lifting of the spin degeneracy of the bands due to the $\mathcal{TI}$ symmetry breaking by the field. }
\end{figure}

\begin{figure*}
\includegraphics[width=2.0\columnwidth]{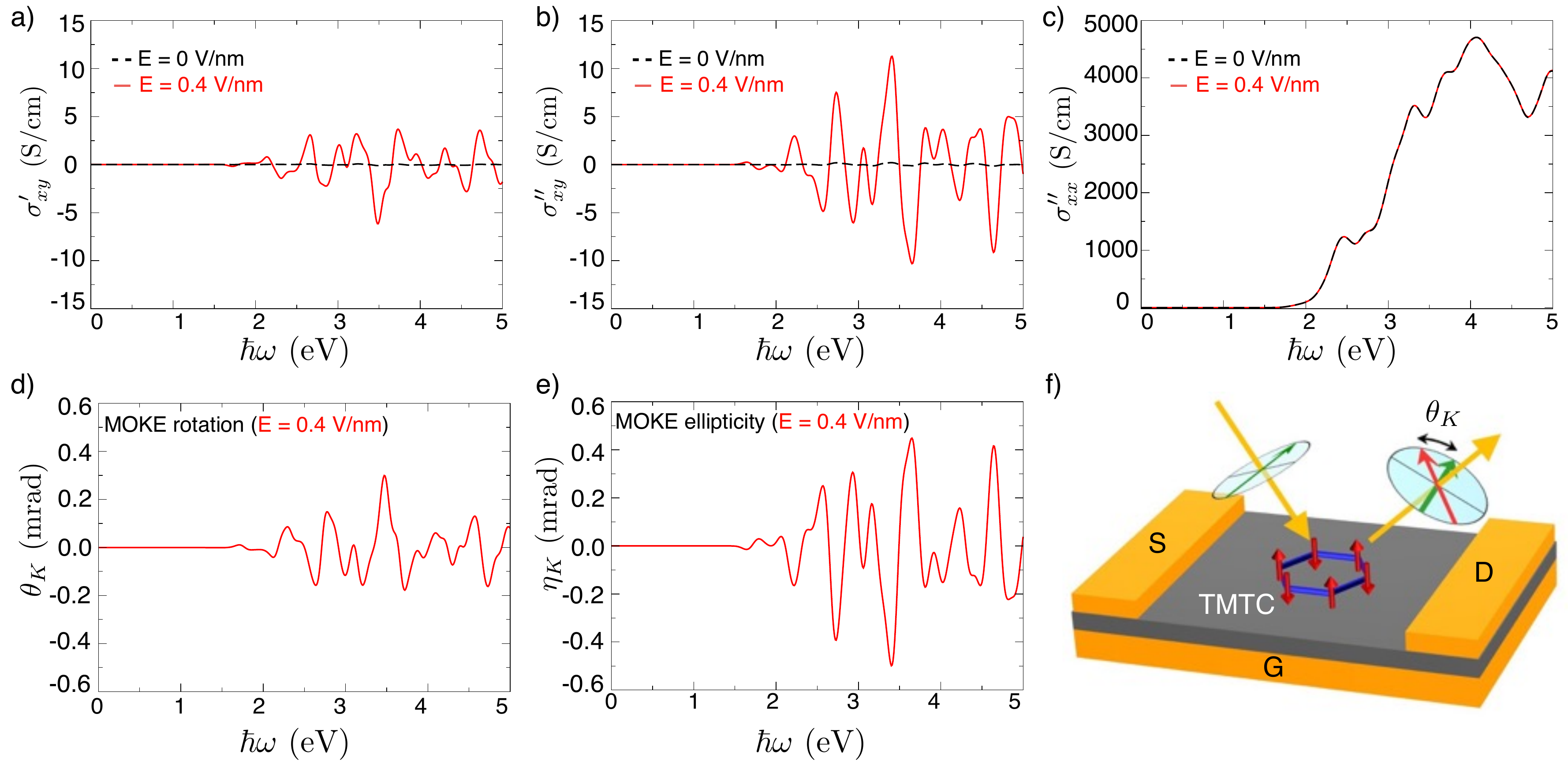}
\caption{\label{fig:Fig3} (Color online) (a) The real part of $\sigma_{xy}$, (b) the imaginary part of $\sigma_{xy}$, and (c) the real part of $\sigma_{xx}$ of bilayer MnPSe$_3$ at zero field (black) and a field with strength 0.4~V/nm (red). The smearing parameter was set to 0.1 eV. The corresponding (d) Kerr rotation angle and (e) ellipticity angle computed as a function of photon energy $\hbar \omega$ for bilayer MnPSe$_3$ on a wedged SiO$_2$ substrate. The zero point of the energy corresponds to the top of valence band. (f) A schematic of a magneto-optic device made from layered antiferromagnets. S, D, and G stand for source, drain and gate respectively. In the incident and the reflected light, an arrow shows the direction of the polarization direction.  On reflection from the antiferromagnets, the plane of polarization of light can be rotated (from green to red arrow), and an ellipticity is induced, depending on the gate voltages.}
\end{figure*}

While crystal symmetries are difficult to control in bulk materials, it has been demonstrated that gating can be an effective tool to break the inversion symmetry in 2D materials~\cite{Zhang09p820,Mak09p256405,Wu13p149,Sui151027,Shimazaki15p1032}.  In the following using first-principles method we demonstrate the idea of gate-controllable MOKE using bilayer MnPSe$_3$ as an example.  In its bulk form, MnPSe$_3$ is a layered compound with weak interlayer Van der Waals interaction. The crystal structure of MnPSe$_3$ monolayer is shown in Fig.~\ref{fig:Fig2}(a).  The magnetic ions (Mn) form a honeycomb lattice within each layer, and each of them is octahedrally coordinated by six Se atoms from its three neighboring (P$_2$Se$_6$) ligands, with the centers of the hexagons occupied by the P$_2$ groups. The Mn ions are in a half-filled $d^5$ state, making MnPSe$_3$ a strong antiferromagnet.  We also find that the system has an easy axis along the $z$-direction, with the spins taking a N{\'e}el-type texture. The bilayer considered here is made of these monolayer units with a stacking order similar to the bulk form [see Fig.~\ref{fig:Fig2}(b)]. There are two Mn atoms in each layer of the bilayer unit cell.  In the top layer, while one Mn atom lies on top of an Mn atom in the bottom layer, the second Mn atom lies on top of the P atoms in bottom layer. The spins of the Mn ions from the two layers are antiferromagnetically coupled. It can be verified that bilayer MnPSe$_3$ has $\mathcal{TI}$ symmetry, hence, no magneto-optic effect is allowed. 

This $\mathcal{TI}$ symmetry can be broken by a perpendicular electric field.  We first look at the effect of such a field on the band structure of bilayer MnPSe$_3$. The details of first-principles calculations are described in Ref.~\cite{sup-MO}.  Figure~\ref{fig:Fig2}(c) shows the band structure in the absence of an electric field.  Because of the presence of the $\mathcal{TI}$ symmetry, the spin-up and spin-down bands are degenerate at each $\bm k$ point, making the material magneto-optically inactive.  However, upon the application of a field (0.4~V/nm), the spin degeneracy of the bands is lifted, symptomatic of $\mathcal{TI}$ symmetry breaking [see Fig.~\ref{fig:Fig2}(d) and its insert]. 
 
Thus, on the application of a perpendicular electric field, we expect bilayer MnPSe$_3$ to become magneto-optically active.  Figure~\ref{fig:Fig3}(a)-(c) show the optical conductivity tensor obtained from the calculation of maximally localized Wannier functions~\cite{PhysRevB.56.12847,Souza01p035109,Mostofi08p685}.  We can see that $\sigma_{xy}$ is zero when the field is zero (black curves).  It becomes non-zero for a finite field (red curves), as expected.  We have also verified that the reversal of the field reverses the sign of $\sigma_{xy}$~\cite{sup-MO}.  The longitudinal conductivity $\sigma'_{xx}$, on the other hand, is almost invariant under the application of a field [see Fig.~\ref{fig:Fig3}(c)].  This is not surprising as $\sigma'_{xx}$ measures the average absorption of right- and left-circularly polarized light~\cite{Feng15p144426}.  We note that the oscillatory behavior of $\sigma_{xy}$ as a function of $\omega$ is already observed in our tight-binding model [see Fig.~\ref{fig:TB}(d)].

To quantify the field-induced MOKE, we have calculated the complex polar Kerr angle.  For simplicity, we assume that the incoming light is perpendicular to the surface, and the sample is placed on a wedged substrate such that there is no reflection from the substrate in the perpendicular direction.  In the thin film limit the Kerr angles are given by~\cite{Kim07p214416,Aguilar12p087403}
\begin{equation} \label{kerr}
\theta_K + i\eta_K  =  \frac{2(Z_0 d \sigma_{xy})}{1-(n_s+  Z_0 d \sigma_{xx})^2}  \;,
\end{equation}
where $\theta_K$ specifies the rotation angle of the major axis of the linearly polarized light, $\eta_K$ specifies the ratio of the minor to the major axis of the light, $n_s$ is the refractive index of the substrate, $Z_0$ is the impedance of free space and $d$ the thickness of bilayer MnPSe$_3$ (10.3~\AA).  Figure~\ref{fig:Fig3}(d) and (e) show the computed MOKE angles for a wedged SiO$_2$ substrate ($n_s =1.5$).  For field strength of 0.4~V/nm, $\theta_K$ can reach up to 0.3~mrad, which is well within the current detection limit~\cite{Kato04p1910,Lee2016p421}.  Note that due to the oscillatory behavior of $\sigma_{xy}(\omega)$, the size of the Kerr angle has a strong dependence on the smearing parameter, and can be made larger in high-quality samples~\cite{sup-MO}.  The smearing parameter $\eta = 0.1$ eV chosen here corresponds to a carrier relaxation time of 6.5~fs, which is in the realistic range for layered transition metal chalcogenides~\cite{MoS2}.   The generation of the MOKE in a magneto-optically inactive material using gate voltage is an important distinction from previous work~\cite{Feng15p144426}. 


We have also studied the field-dependence of the MOKE in monolayer MnPSe$_3$. Similar to bilayers, monolayer MnPSe$_3$ also has $\mathcal{TI}$ symmetry.  However, we find that the MOKE angle remains negligibly small in monolayers upon the application of an  electric field of the same strength~\cite{sup-MO}.  This is due to the fact that in monolayer MnPSe$_3$, the inversion symmetry breaking is realized by creating a potential difference between the top and bottom PSe$_3$ layers, which is ``felt'' by the Mn atom through the interaction between the Mn $\textit{d}$ orbitals and the Se $\textit{p}$ orbitals.  This is a much weaker effect compared to the case of bilayers where the Mn atoms in different layers directly feel the effect of the electric field. 

Our predicted gate-controllable MOKE has important implications in both fundamental research and practical applications.  As the observed MOKE is very sensitive to the underlying magnetic order, it can be used to identify the magnetic ground state.  Not only can this method distinguish between ferromagnets and antiferromagnets, but it can be also used to distinguish among different antiferromagnetic orders, such as N{\'e}el, zigzag and stripy order on a honeycomb lattice~\cite{Sivadas15p235425}, supplemented by symmetry analysis and band structure calculations.  This is especially valuable for 2D materials since neutron scattering is ineffective for these materials due to the small scattering cross section.  Furthermore, the sensitivity of the MOKE to the magnetic order can be exploited for magnetic information storage.  For instance, the reversal of the N{\'e}el vector will result in a change of sign of the observed MOKE. Thus, the information encoded in the N{\'e}el vector can be extracted using this gate-controlled MOKE in antiferromagnets. 

We are grateful to Hua Chen, Matthew W.~Daniels, Tony Heinz, Kin Fai Mak, David Mandrus, Jiaqiang Yan, and Xiaodong Xu for stimulating discussions. We would also like to thank Valentino Cooper, Ji Feng and Xiao Li for their computational inputs.  We are indebted to the anonymous reviewers for providing insightful comments on an earlier version of this work.  This work was supported by the Air Force Office of Scientific Research under Grant No.~FA9550-12-1-0479 and FA9550-14-1-0277, and by the National Science Foundation under Grant No.~EFRI-1433496.  S.O.\ acknowledges support by the U.S.\ Department of Energy, Office of Science, Basic Energy Sciences, Materials Sciences and Engineering Division. This research used resources of the National Energy Research Scientic Computing Center, which is supported by the DOE Office of Science under Contract No.~DE-AC02-05CH11231.  D.X. also acknowledges support from a Research Corporation for Science Advancement Cottrell Scholar Award.

%

\end{document}